\documentclass[pra,twocolumn,notitlepage,showpacs]{revtex4-1}
\usepackage[usenames,dvipsnames]{color} 
\usepackage{graphicx,amssymb,amsmath}
\newcommand{\sx}{\sigma_x}
\newcommand{\sy}{\sigma_y}
\newcommand{\sz}{\sigma_z}
\newcommand{\ham}{{\mathcal{H}}}
\newcommand{\tr}{\mathrm{Tr}}
\DeclareMathSymbol{\theta}{\mathalpha}{letters}{"23}
\definecolor{LinkColor}{rgb}{0,0,.5}
\renewcommand{\emph}{\textit}
\usepackage[colorlinks=true,citecolor=LinkColor,urlcolor=LinkColor]{hyperref}

\begin{document}
\title{Time-optimal control with finite bandwidth}
\author{M. Hirose and P. Cappellaro}
\affiliation{Department of Nuclear Science and Engineering and Research Laboratory of Electronics, Massachusetts Institute of Technology, Cambridge, MA, USA}

\begin{abstract}
Optimal control theory provides recipes to achieve quantum operations with high fidelity and speed, as required in quantum technologies such as quantum sensing and computation. While technical advances have achieved the  ultrastrong driving regime  in many physical systems, these capabilities have yet to be fully exploited for the precise control of quantum systems, as other limitations, such as the generation of higher harmonics or the finite bandwidth of the control fields, prevent the implementation of theoretical time-optimal control. Here we present a method to achieve time-optimal control of qubit systems that can  take advantage of  fast driving beyond the rotating wave approximation. We exploit results from optimal control theory to design driving protocols that can be implemented with realistic, finite-bandwidth control and we find a relationship between  bandwidth limitations and achievable control fidelity. 
\end{abstract}

\maketitle
Precise control of quantum systems is a requirement for many applications of quantum physics, from quantum information processing to quantum metrology and simulation. Fast control is highly desirable to beat decoherence and improve performance of these quantum devices. This desire has spurred much research on the ultimate control speed for unitary~\cite{Deffner13,Salamon09} and dissipative~\cite{DelCampo13} dynamics, as well as shortcuts to adiabatic control~\cite{Bason12}.
At the same time, technological advances have enabled driving quantum transitions faster than the natural transition frequency, in  systems ranging from atoms~\cite{Hofferberth07,Jimenez-Garcia15} to quantum wells~\cite{Zaks11}, from superconducting qubits~\cite{Deng15,Ashhab07,Rudner08,Oliver09,Niemczyk10} to mechanical oscillators~\cite{Barfuss15} and isolated spin defects~\cite{Fuchs09,Childress10,Scheuer14}. In this ultrastrong driving regime, the design of control protocols can no longer rely on the usual intuition, based on the rotating wave approximation (RWA). While optimal control theory gives prescriptions to achieve time-optimal (TO) control, often the ideal control schemes cannot be applied in practice, due to bounds in the control strength, phase or bandwidth. 
For example, bounds in the control strength impose a \textit{quantum speed limit}~\cite{Margolus98,Carlini06,Giovannetti03,Caneva09,Hegerfeldt13} on the system evolution, which is related to an energy-time uncertainty relation~\cite{Deffner13,DelCampo13}.
Limitations on the control of the driving field phase or polarization preclude the application of many TO control schemes. For example, it  has been shown~\cite{Boscain02a,DAlessandro01,Albertini15} that for a two-level system (qubit), the TO solution is given by an ``on-resonance'' driving, if   the phase or polarization of the driving field is under experimental control~\cite{London14,Shim14}. 
When this is not possible, the internal Hamiltonian (the \textit{drift} term)  cannot be eliminated and the TO solution takes the form of a bang-bang (BB) control~\cite{Boscain06,Billig13,Aiello15q,Billig14,Avinadav14}. This optimal solution assumes that there are no limitations in the control bandwidth; however, in practice the control fields cannot be switched on and off instantaneously. 
Here we show that we can approach time-optimal driving of a qubit, given a bound, real driving field along a single axis, $|\Omega(t)|\leq\overline\Omega$, even when the Fourier transform of $\Omega(t)$ is defined over a finite range $[0,\Delta\omega]$. 
With the goal of keeping the gate time equal to the BB optimal time, we construct an analytical control strategy based on a Fourier series approximation to the ideal control. Our \textit{Fourier-Approximated  Time-Optimal} (FATO) control strategy achieves several  key results. 
First, it provides an analytical recipe to design high fidelity, time-optimal control sequences in the regime of ultrastrong driving, when the RWA breaks down. Even in the case of weak driving, where the RWA is applicable, it achieves shorter gate times than  conventional (on-resonance) methods. 
Just as importantly, we identify control bandwidth as a limiting resource in the compromise between fidelity and time-optimality~\cite{Moore12,Lloyd14l}. Even if a Fourier approximant is not the only solution, it allows us to easily analyze bounds on the control fidelity that bandwidth limitations impose, with analytical solutions describing the dependence of gate fidelity on the bandwidth. 
In addition, we show that the FATO scheme is robust against errors in the control field and it can further be extended to  controlling more than one qubit.

\begin{figure}[b]\centering
\includegraphics[width=0.13\textwidth]{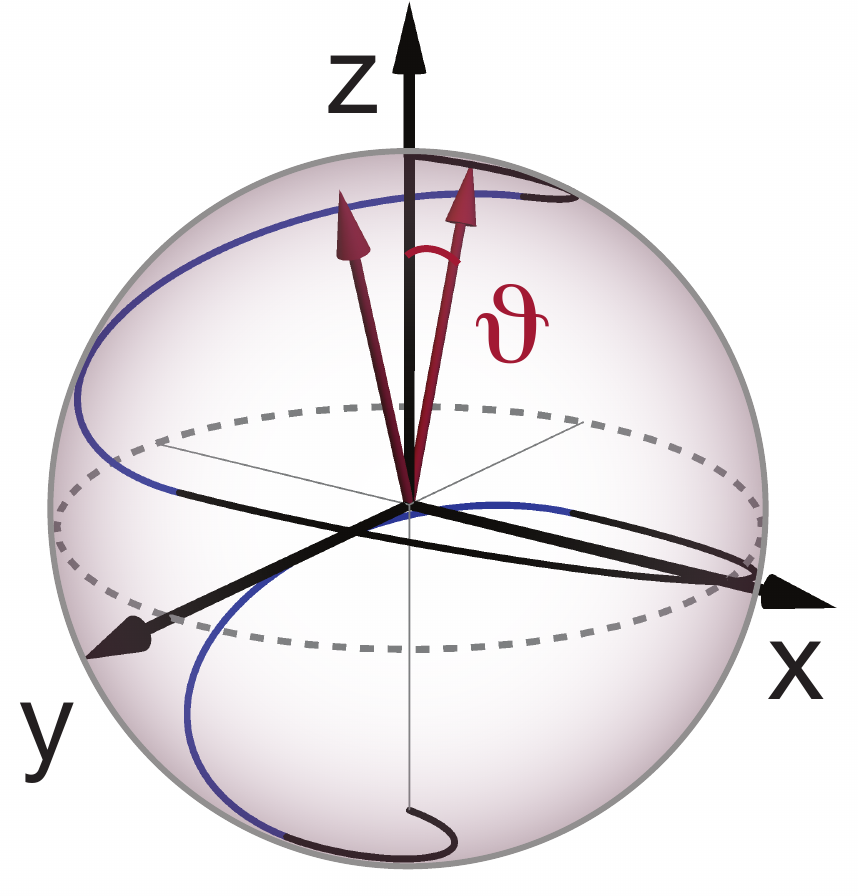}
\   \includegraphics[width=0.33\textwidth]{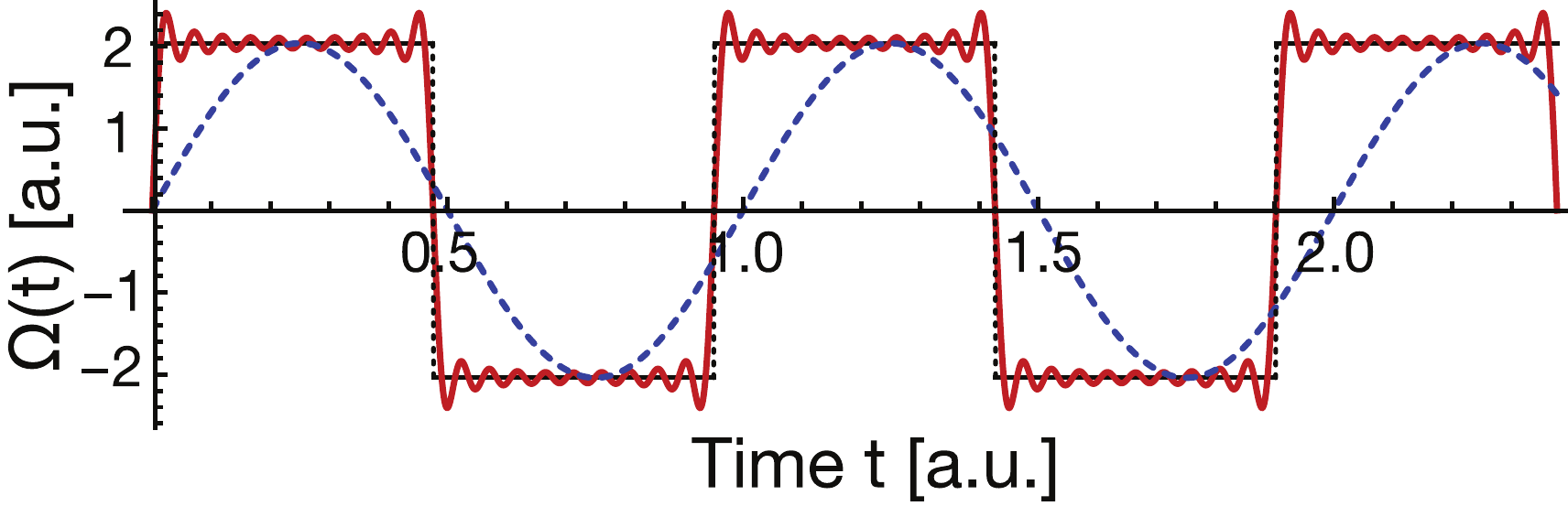}  
\caption{
\textit{Left: Time-optimal control on the Bloch sphere}. The figure shows the two axes of rotation (red arrows) separated by the angle $2\theta\!=\!\pi/5$. We plot a representative trajectory (blue-black line) from $|0\rangle$ to $|1\rangle$ achieved  with the TO BB control shown on the right. \textit{Right: FATO waveform}. Here we considered the BB solution (solid black line) to achieve a $\pi$ rotation about X, for $\theta\!=\!\pi/10$ and $\omega_0\!=\!\pi$ and its Fourier Series (red solid line) truncated at $k\!=\!57$ (9 non-zero coefficients). The dashed line is the waveform for on-resonance driving (notice that its period does not match the BB period).}  
  \label{fig:Waveform}
\end{figure}

\textbf{Fourier-Approximated  Time-Optimal Control --}
Assume we have a qubit with internal (\textit{drift}) Hamiltonian $\ham_0=\omega_0\sz/2$ and we can apply a control  $\ham_c=\Omega(t)\sx/2$, with $\Omega$ real and bounded by $|\Omega(t)|\leq\overline\Omega$. This situation is relevant to many experimental systems, from nuclear and electronic spin resonance to atomic systems and superconducting qubits. The control Hamiltonian is generated by,  e.g., radiofrequency or microwave fields applied along one physical axis by a wire or antenna in the experimental setup, which can set the time-dependent amplitude of the field source.
The goal is to perform a desired unitary evolution in a time-optimal way.

The usual strategy to achieve precise qubit control  is to rely on the rotating wave approximation: for example, to achieve a rotation about $x$, we set $\Omega_{\mathrm{RWA}}\!\ll\!\omega_0$ and drive on resonance, $\Omega(t)=\Omega_{\mathrm{RWA}}\cos(\omega_0t)$. 
More general rotations can  be obtained by choosing the frequency and phase of the driving, thus making it possible, for example, to effectively drive along the perpendicular direction ($y$-axis) even if the driving field is along the laboratory $x$-axis. 
This solution is however not time-optimal: indeed, one effectively only uses half of the driving strength, as the other half is the counter-rotating field.
 More precisely, Pontryagin's minimum principle~\cite{Pontryagin87b} can be used to prove that for this control problem a BB sequence with $\Omega(t)=\{\pm\overline\Omega,0\}$ is the TO solution~\cite{Boscain06,Garon13}. In addition, if experimental conditions allow $\Omega\gtrsim\omega_0$, the RWA is not applicable and on-resonance driving is no longer a good control strategy.

For both cases (ultrastrong or weak driving) the ideal TO control strategy for one-axis driving is to always evolve at the maximum ``speed'' (BB control), assuming one can switch the sign of the function $f(t)=\Omega(t)/\overline\Omega$ infinitely fast. The total Hamiltonian then takes the form
\begin{equation}\renewcommand*{\arraystretch}{1.5}\vspace{-4pt}
\begin{array}{l}
\ham_\pm=\frac12[\omega_0\sz\pm\overline\Omega\sx]=\frac\omega2(\sz\cos\theta\pm\sx\sin\theta),\\ \ham_0=\frac12\omega_0\sz, \end{array}\vspace{-8pt}
\label{eq:Ham}
\end{equation}
where we defined $\omega=\sqrt{\omega_0^2+\overline\Omega^2}$ and $\tan(\theta)=\overline\Omega/\omega_0$. 
The control is obtained by switching between these three Hamiltonians inducing rotations about three different axes.  The SU(2) gate synthesis problem then reduces to finding the times $t_j^{(+,0,-)}$ for each ``bang''. 
Strong conditions on these times and number of bangs have been recently found using an algebraic solution~\cite{Aiello15q,Billig13,Billig14}. Only three parameters are necessary to describe the TO solution~\cite{Billig13}, an initial and final time $t_i$ and $t_f$, while middle times  $t^\pm_m$ are related to each other. In particular,  when the rotation speed about  the two axes is equal,  which is the case here, the middle times are all equal. 
Properties of the TO decomposition can be classified based on the angle $2\theta$ between the two rotation axes, and in particular whether $\theta \lessgtr \pi/4$, corresponding to weak ($\overline\Omega<\omega_0$) or strong ($\overline\Omega>\omega_0$) driving. For example, for $\theta>\pi/4$ finite solutions have at most 4 bangs, while for $\theta<\pi/4$ they can have an increasing number of bangs with decreasing $\theta$~\footnote{Infinite decompositions are also possible. However, here an
  infinite sequence with equal times is a rotation about $\sigma _z$, speedily obtained by singular control
  ($\Omega =0$).}.
These constraints (see also Appendix~\ref{sec:appendix}) 
can be used to efficiently search for TO solutions to the synthesis of any desired unitary in SU(2), under the assumption of \textit{infinite bandwidth} of the control function (infinitely fast switching between $\pm\overline\Omega$). 
In the following we show how, even when the control bandwidth is limited, the BB solution  forms the basis for an excellent TO control scheme that we call \textit{Fourier-Approximated Time-Optimal} (FATO) control.

We assume that the control field can only be switched with a finite speed; this sets an upper bound $\Delta\omega$ to its bandwidth.
The ideal TO solution, with total sequence time $T$ and switching times $\{t_i,t_m,t_f\}$, effectively defines a piecewise-constant function $f(t)=\Omega(t)/\overline\Omega=\{\pm1,0\}$. It is always possible to express $f(t)$ over the interval $[0,T]$ as a Fourier series (see Fig.~\ref{fig:Waveform}): 
\begin{equation}
f(t)=\frac{c_0}2+\sum_{k=0}^\infty[s_k\sin(2\pi k t/T)+ c_k\cos(2\pi k t/T)].
\label{eq:fourier}
\end{equation}
 Then, imposing the bandwidth constraint on the control function is equivalent to truncating this sum to $k=K$ with $2\pi K/T\leq \Delta\omega$. In turns, this will reduce the fidelity of the control, while preserving its duration at the optimum time. 

\textbf{Fidelity and Robustness of FATO Control --}
\begin{figure}[t]\centering
  \includegraphics[scale=0.33]{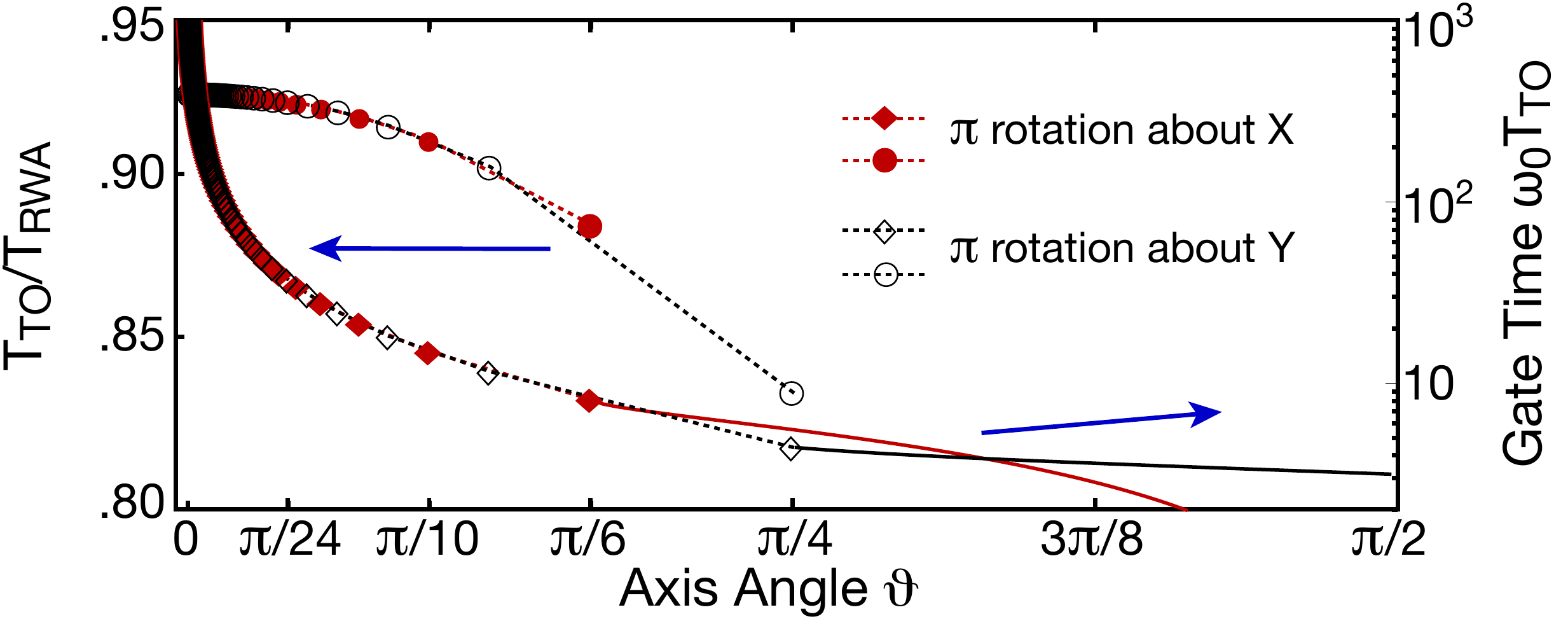}
  \caption{\textit{Left axis:} Comparison between the time required for a $\pi$ rotation with the FATO  method and the RWA on-resonance driving as a function of driving strength (parametrized by the angle $\theta$). Note the comparison is only possible for $\theta\leq\pi/4$. \textit{Right axis:} Normalized time $\omega_0T$ required for $\pi$ rotations. Circles (and dotted lines) are for the weak driving solution, while solid lines are for the ultrastrong driving regime.}
  \label{fig:TimeComp}
\end{figure}
For a given bandwidth $\Delta\omega$, two properties of the time-optimal BB solution will determine how well it can be approximated by FATO: the minimum non-zero time among $\{t_i,t_f,t_m\}$ (shorter times requiring larger bandwidths) and the number of switches (larger $n$ requiring in general larger $k$ for a better approximation).
In general, the middle times are constrained by $t_m\geq\pi/\omega=\pi\cos(\theta)/\omega_0$; they define a square wave with $n\leq \lfloor \frac\pi\alpha \rfloor +1$ switches. The minimum bandwidth is then $\Delta\omega\geq \sqrt{\omega_0^2+\overline\Omega^2}= \omega_0/\cos(\theta)$, that is, it depends not only on the ``resonant'' frequency $\omega_0$, but also on the driving strength, as stronger driving allow for faster control, thus requiring larger bandwidth for time optimality.

To make our method more concrete, we focus on exemplar target gates,  $\pi$ rotations about the $x$- and $y$-axis. 
These gates are particularly important (they are ``NOT'' quantum gates) and describe an evolution under the control operator only, eliminating the effects of the drift.  
While focusing on these gates allows us to find explicit solutions for the BB TO problem that are the starting point for the bandwidth-limited construction, the same analysis would also apply to other unitaries. 
For these gates we can more easily analyze the performance of FATO control in terms of gate time, fidelity as a function of bandwidth and robustness to imperfections. 

\begin{figure}[t]\centering
  \includegraphics[scale=0.5]{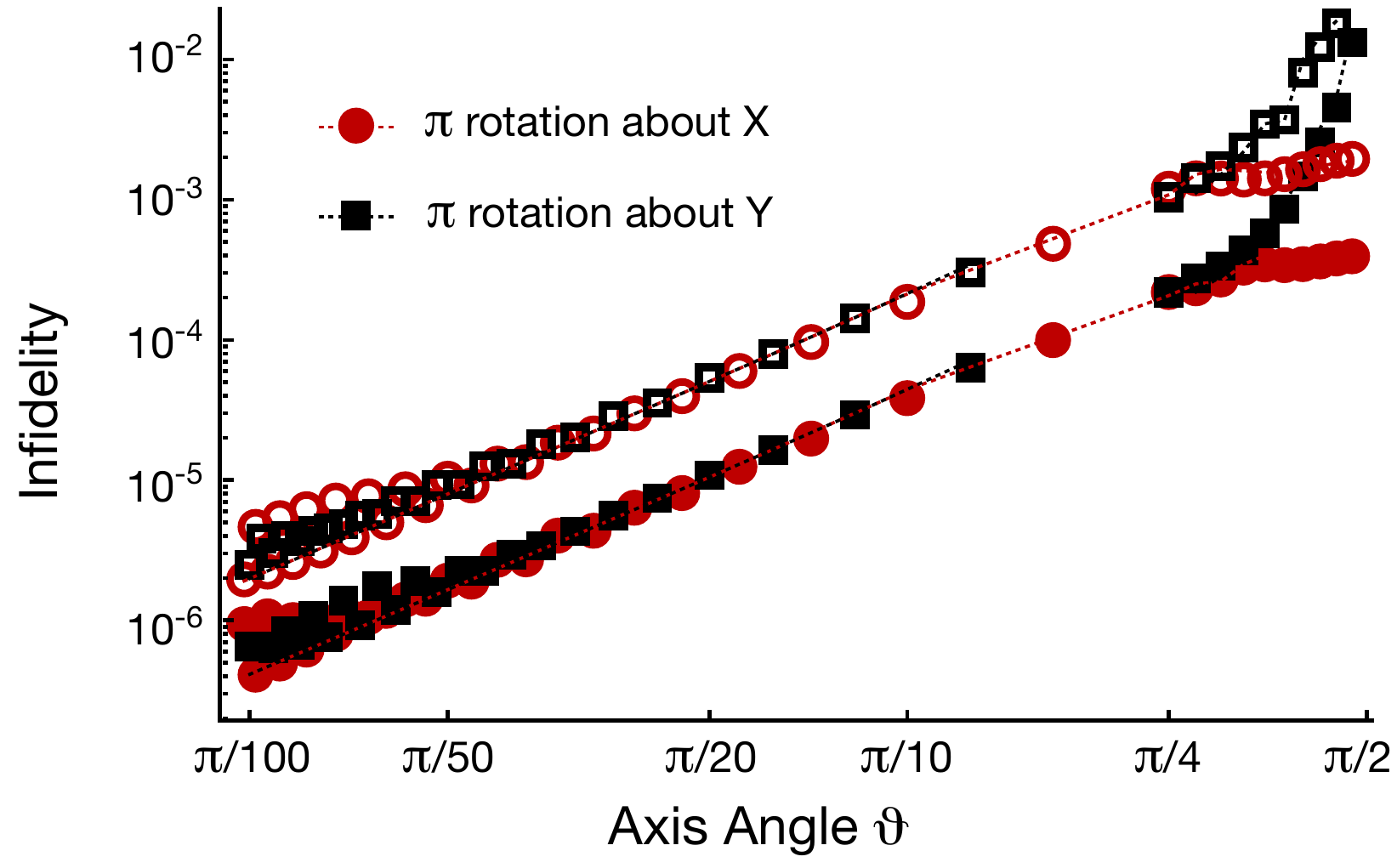}
  \caption{Infidelity for a $\pi$-pulse about X (red) and Y (black) as a function of the driving strength, parametrized by the angle  $\theta$. Open symbols are for a Fourier expansion with 5 non-zero coefficients and filled symbols with 19 non-zero coefficients (the resulting bandwidth depends on the angle $\theta$). The dashed lines are the analytical expressions (not a fit) based on the Fourier series approximation error $\mathcal E_K$.}
  \label{fig:FidTheta}
\end{figure}
\textit{Gate Time --}
We distinguish between weak and ultrastrong driving, as they have different BB solutions.
In the case of weak driving, the angle between the two axes of rotation is small and we expect generally longer control sequences (large $T$) with many bangs (large $n$). While specific solutions for arbitrary $\theta$ must be found numerically, analytical solutions are available for specific values. In particular, we find that the optimal solution has  $n=\frac{\pi}{2\theta}$ bangs for X(Y) $\pi$ rotations, whenever $n$ is an odd(even) integer number (see Appendix~\ref{sec:appendix}). All the bang times are equal and such that $\omega t_m=\pi$. The function $f(t)$ is then a simple square wave with period $2\pi/\omega$. 
The total time is $T_{\mathrm{TO}}=n\pi/\omega=\pi^2\cos(\theta)/(2\theta\omega_0)$. In Figure~(\ref{fig:TimeComp}) we compare this optimal time to the time required with on-resonance driving, $T_{\mathrm{RWA}}=2\pi/\overline\Omega$ (as the effective Rabi frequency in the RWA is $\overline\Omega/2$). The ratio of the two strategy times is given by $T_{\mathrm{TO}}/T_{\mathrm{RWA}}=\frac{\mathrm{Si}(\pi)\sin\theta}{2\theta}$ (with $\mathrm{Si}$ the sine integral function), where we took into account that due to Gibbs phenomenon~\cite{Gibbs98,Bocher06}, the Fourier series approximation yields an effective larger maximum driving frequency, $\Omega'\approx \frac{2 \mathrm{Si}(\pi )}{\pi }\overline\Omega$. 

In the ultrastrong driving regime a direct comparison with the time required for on-resonance driving  is not possible, since the RWA is violated. Our method still  provides  a constructive strategy to achieve control beyond the RWA, and does so in a time-optimal way.
The TO solution for strong driving consists of $n=3$ bangs, with the middle bang  singular ($\Omega=0$) to obtain a $\pi$ pulse about Y. The times are given by
\begin{equation}
\begin{array}{ll}
t_1^x\!=\!t_3^x\!=\!\frac{2  \mathrm{arccsc}[2 \sin(\theta )]}{\omega },&  t_1^y\!=\!t_3^y\!=\!\frac{2 \mathrm{arccot}\left[\sqrt{\!-\!\cos (2 \theta )}\right]}{\omega } \\
t_2^x\!=\!2\pi\!-\!\frac{2  \mathrm{arccsc}[2 \sin(\theta )]}{\omega },& t_2^y\!=\!\frac{2 \arctan\left[\sqrt{\tan(\theta)^2-1}\right]}{\omega_0}
\end{array}
\label{eq:TimeS}
\end{equation}
These times define the piece-wise constant function $f(t)$ that we approximate with a Fourier series expansion to obtain the FATO control driving field shape. 
\begin{figure}[b]\centering
  \includegraphics[scale=0.33]{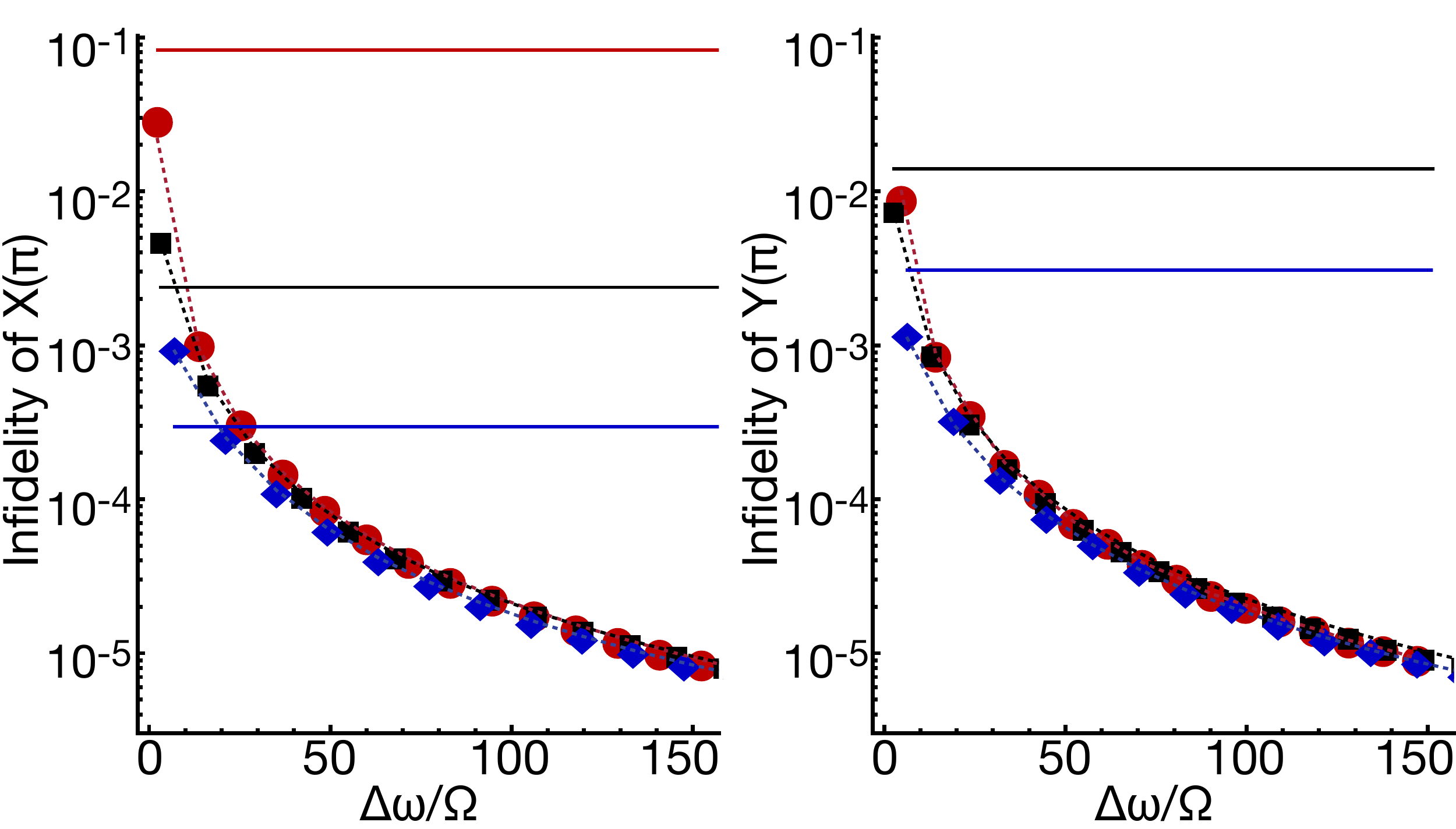}
  \caption{Infidelity for a $\pi$-pulse about X (left) and Y (right) as a function of the bandwidth $\Delta\omega/\Omega$. We consider ultrastrong driving, $\theta\!=\!\pi/3$ (red), and weak driving, $\theta\!=\!\pi/10, \pi/8$ (black) and $\theta\!=\!\pi/22, \pi/20$ (blue) for X and Y respectively. The dotted lines are the infidelities (not a fit) calculated from the Fourier series error $\mathcal E_K$. The solid lines are the infidelities (due to breaking of the RWA) of the on-resonance driving strategy (the infidelity of Y($\pi$) for $\pi/3$ is not shown as it is $>.1$).}
  \label{fig:FidDrive}
\end{figure}

\textit{Fidelity -- }
Because of the finite bandwidth of the control field, the ideal gate propagator cannot be perfectly implemented in the optimal time. In our FATO approach, we keep fixed the gate time at its optimum value; we can then evaluate the effects of the finite bandwidth by calculating the entanglement fidelity~\cite{Nielsen02}. Truncating the Fourier expansion to order $K$ implements the propagator $U_K$ satisfying 
\begin{equation}
i\dot U_K(t)\!=\!\frac{\omega_0}2[\sz+\tan(\theta)[f(t)- R_K(t)]\sx]U_K(t),
\label{eq:UK}
\end{equation}
where we defined the reminder of the truncated Fourier series of $f(t)$,
\begin{equation}
R_K(t)=\sum_{k=K+1}^\infty[s_k\sin({2\pi k t}/T)+ c_k\cos({2\pi k t}/T)].
\label{eq:reminder}
\end{equation}
The fidelity $F\!=\!|\tr[U_{id}^\dag(T)U_K(T)]|/2$ with respect to the ideal propagator, $U_{id}(T)$, can be computed in terms of a reminder propagator, $U_R(t)=U_{id}^\dag(t)U_K(t)$.
$U_R(t)$  can be evaluated by moving to the ideal Hamiltonian  \textit{toggling} frame, where the Hamiltonian is
$\widetilde\ham(t) = U^\dag_{id}(t)[\ham(t)-\ham_{id}(t)]U_{id}(t)$. Assuming the reminder is small, we can  approximate $\widetilde U_R$ with a first-order Magnus expansion given by the effective Hamiltonian 
$\overline\ham_R=\int_0^T \widetilde\ham(t')dt'$. In the weak coupling regime, for example, we obtain 
\begin{equation}
\overline\ham_R=\frac{\omega_0\theta}\pi\tan(\theta)\mathcal E_K[\sx-\tan(\theta)\sz],
\end{equation}
where we introduced the mean error of the truncated Fourier series (see Appendix~\ref{sec:appendixB}):
\begin{equation}
\mathcal E_K= \frac{2}{T}\int _0^T \!\!R_K^2(t)dt=\frac12\sum_{k=K+1}^\infty (c_k^2+s_k^2).\end{equation}
This yields the fidelities $F_w^{x,y}\approx\cos[\tan(\theta)\mathcal E_K \pi/4]$.  While in the strong driving regime the exact calculation is more complex, we still find that the fidelities  are well approximated by a function of the mean errors, $F_s^x\approx\cos[2/\pi\sin(\theta)\mathcal E_K]$ and $F_s^y\approx\cos[2/\pi\tan(\theta)\mathcal E_K]$, as shown in Fig.~(\ref{fig:FidTheta}) and (\ref{fig:FidDrive}).
\begin{figure}[t]\centering\qquad
    \includegraphics[scale=0.33]{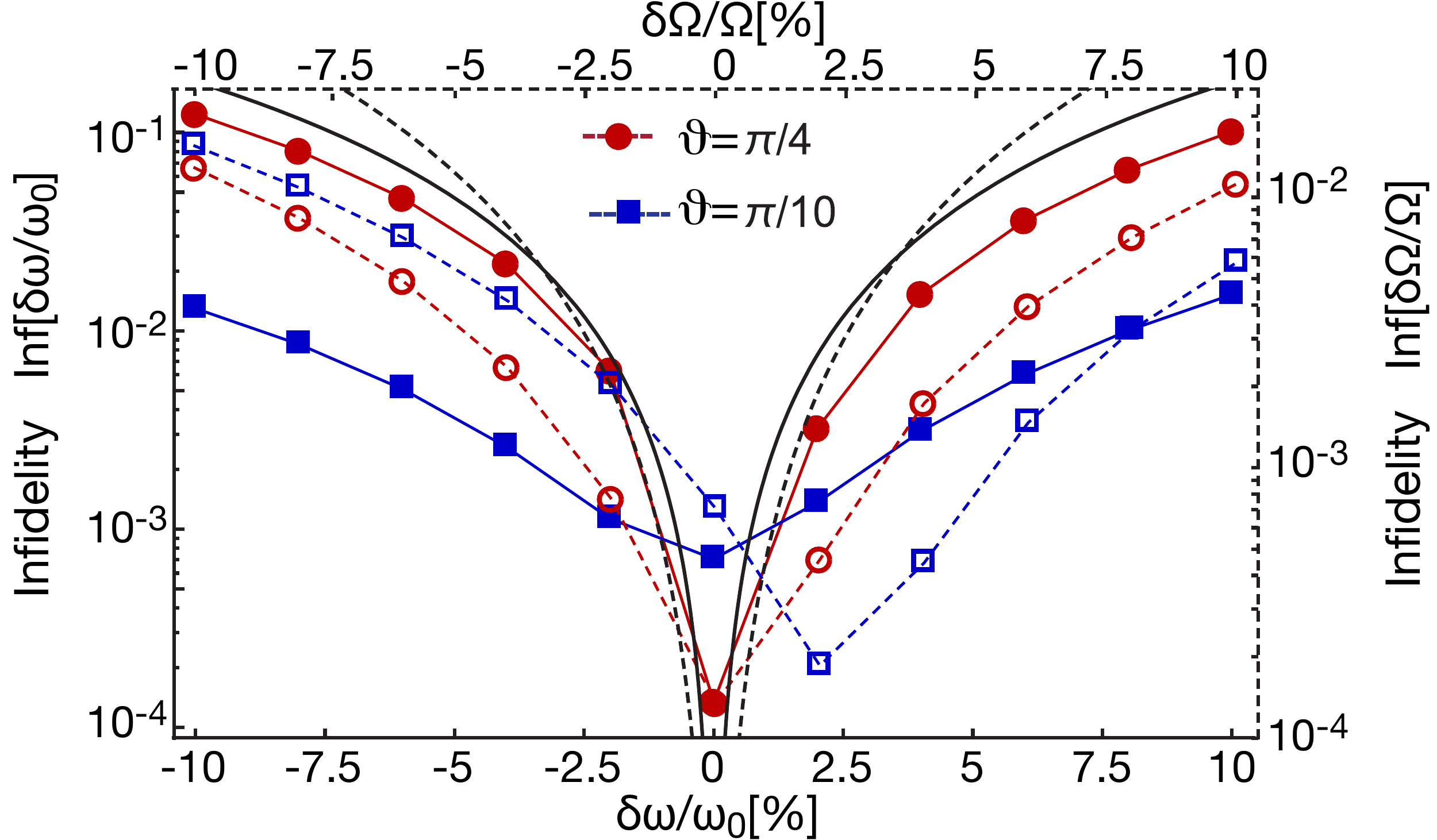}
  \caption{Infidelity for a $\pi$-pulse about X   as a function of the error in frequency $\omega_0$ (filled symbols, left axis) or driving strength $\Omega$ (open symbols, dashed line, right axis).  We consider two exemplary cases for strong ($\theta=\pi/4$, red) and weak ($\theta=\pi/10$, blue) driving.
Although the fidelity of the gates decreases with the inhomogeneities, the error is typically smaller than for the usual on-resonance driving.}
  \label{fig:Inhomogeneities}
\end{figure}

We thus found a simple relationship between the fidelity and 
the Fourier series mean error, $\mathcal E_K$, which encompasses the Fourier properties of the TO BB function and the available control bandwidth. This relationship not only  makes it  possible to easily find the required bandwidth for a desired fidelity, but also provides insight onto which BB controls require larger bandwidth. For example, in the ultrastrong regime, as $\overline\Omega/\omega_0\!\to\!\infty$ ($\theta\!\!\to\!\pi/2$), the times required for an X rotation go to zero, thus allowing good fidelity; however, a Y rotation still requires a finite time, reducing the fidelity for the same bandwidth (Fig.~\ref{fig:FidTheta}).

Still, as shown in figure (\ref{fig:FidDrive}), the infidelity, $\textrm{Inf}=1\!-\!F,$ decreases rapidly with the control bandwidth. In addition, in the weak driving regime, FATO control beats the fidelity obtained with the on-resonance driving (when taking into account the counter-rotating field), even when considering a very small bandwidth ($\Delta\omega\approx \omega_0\div2\omega_0$). Very good fidelity is also obtained in the strong driving regime. We remark that since very high bandwidth can be routinely reached in experiments, our construction could  achieve fidelity beyond the fault-tolerance threshold~\cite{Gottesman98}, while still operating at the maximum quantum speed.

\textit{Robustness to parameter variations --}
We further evaluate the robustness of the FATO control scheme with respect to errors in either the frequency $\omega_0$ or the driving strength $\overline\Omega$. The first case corresponds to the situation where either the internal Hamiltonian is not known with sufficient precision or there are variations due to inhomogeneities. In the second case, we analyze the possibility of an error or inhomogeneity in the driving power.
Typical results are shown in Fig.~(\ref{fig:Inhomogeneities}). We find that even for a few percent error, the fidelity of the gate is good and it is typically higher than for the usual on-resonance driving.
In some cases, the fidelity can be even higher than in the absence of error: this is due to the fact the error can contribute to either driving fields or larger bandwidth. We note however that in these cases the driving in the presence of errors might not be anymore time-optimal.
\begin{figure}[t]\centering
    \includegraphics[scale=0.33]{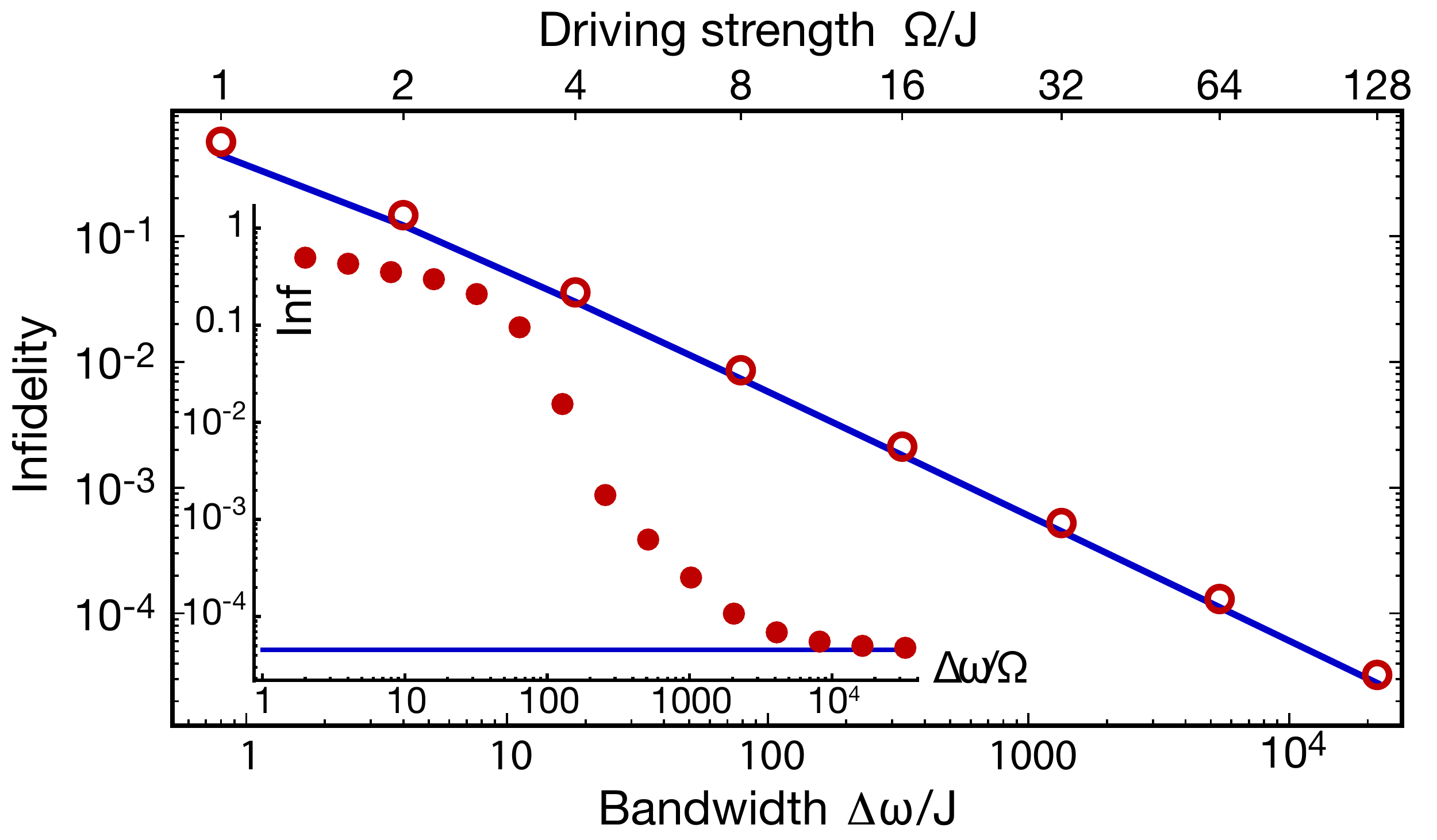}
  \caption{Infidelity for a SWAP gate on two qubits   as a function of the bandwidth $\omega/J$ for varying driving strengths $\Omega/J$. The blue line is the fidelity of the ideal BB control (with finite pulses) and the red dots its FATO approximation. In the inset: Infidelity as a function of bandwidth for fixed driving strength, $\Omega=100J$.}
  \label{fig:Fid2qb}
\end{figure}

\textbf{Extension to two-qubit systems --} Until now we focused on TO control of a single qubit. Our strategy can be however extended to larger systems with the combined goals of finding TO control laws and their fidelity dependence on available bandwidth. For example, the BB solutions we found to generate NOT gates for a single qubit could also be used to drive two qubits with opposite internal Hamiltonian~\cite{Romano15}. Then, all the results found here on the effects of a limited bandwidth still apply. 

We can further analyze time-optimal sequences that have been proposed to generate two-qubit gates~\cite{Khaneja01} under the assumption of delta pulses. Using a Cartan decomposition, it was found that TO control of two qubits can be achieve with singular BB control. Adopting the TO solution and introducing finite-length pulses reduces the fidelity; assuming a finite bandwidth (so that ideal, rectangular pulses cannot be applied) further degrades the fidelity. We can use the FATO construction to evaluate these limitations.  We consider for example a quantum SWAP gate (see Appendix~\ref{sec:appendix}, given by three free-evolution periods under the Hamiltonian $H_0=J\sz\sz/2$ interrupted by $\pi/2$-pulses about $x$ and $y$. Assuming a strength $\Omega$ of the driving fields, we plot in Fig.~(\ref{fig:Fid2qb}) the FATO fidelity as a function of bandwidth, demonstrating the good performance of our method.

\textit{In conclusion}, we devised a strategy to drive qubits in a time-optimal way, with high fidelities limited only by the available bandwidth of the control. The technique is in particular useful for ultrastrong driving fields, where intuitive approaches based on the rotating wave approximation fail and only numerical approaches were available until now~\cite{Motzoi11,Bartels13,Caneva11,Doria11}. In addition, our analytical solution provides a simple way to evaluate the required control bandwidth for a desired fidelity. The principles of FATO design could be further extended to achieve control in larger systems, for example to achieve the simultaneous time-optimal control of many qubits~\cite{Assemat10,Burgarth10,Romano14,Romano15} or two-qubit gates~\cite{Khaneja01}.
While our method already provides high-fidelity control, it could be further used as an initial guess for numerical searches~\cite{Doria11} if higher fidelity is desired or to achieve control of 1-2 qubits embedded in larger systems. 
As ultrastrong driving becomes attainable in a number of solid-state and atomic quantum systems, from superconducting qubits to isolated spins, our control strategy will enable taking full advantage of these technical capabilities to manipulate qubits in a time-optimal way. Beyond providing a recipe for TO control, our construction also allows us to explore the compromise between fidelity and time-optimality, by linking gate errors to the available control bandwidth.

%

\appendix
\section{Time-Optimal Bang-Bang control}\label{sec:appendix}
Bang-bang control has been shown to achieve time-optimal control of two-level systems. General  bounds and prescriptions for the time-optimal bang-bang control have been provided~\cite{Billig13,Boscain05,Boscain06,Boscain14,Romano15,Albertini15,Aiello15q}. These results can be used to numerically obtain  a solution to the time optimal problem with BB control for a general unitary. 
However, for some target unitaries and Hamiltonian parameters it is possible to find analytical solutions. In the main text we focused on these cases since they allow to more easily study trends in the fidelity and robustness of the FATO control strategy. Here we describe how, exploiting known results in BB control, we obtained the specific TO solutions  for the two gates and two driving strength regimes considered.

The general goal is to find the optimal times such that a sequence of ``bangs'' under alternating Hamiltonian $\ham_{\pm,0}$ can achieve the desired unitary.
Simple algebraic arguments~\cite{Billig13} impose constraints on the middle times of any TO decomposition. 
Then the TO problem reduces to finding three times, $t_i, t_m$ and $t_f$, such that concatenating the unitaries $U_{0,\pm}(t)=e^{-it\ham_{0,\pm}}$ achieves the two desired gates, $\sx$ and $\sy$. 
Since we focused on achieving $\pi$ rotations, we can apply (in addition to results found in \cite{Billig13,Billig14,Aiello15q}) the results of \cite{Boscain06} relating to a north-to-south pole transformation only.  Their constraints still need to be valid sufficient conditions (although not necessary) for the TO unitary.\vspace{-12pt}
\subsection{Weak Driving}
For weak driving, it was found in~\cite{Boscain06} that there should be no singular bangs in the TO solution. In addition, for $\alpha=\pi/2n$ the solution $U_{NS}$ for the north-to-south transition is obtained as 
\begin{equation}
U_{NS}=[U_+(\pi)U_-(\pi)]^n
\end{equation}
We find that if $n$ is even, $U_{NS}=\sy$, while if $n$ is odd $U_{NS}=\sx$. Ref.~\cite{Boscain06} allowed for a second type of solution, with $n+1$ bangs. These are candidates solutions for $\sy$ if $n$ is odd and $\sx$ if $n$ is even. While it is possible to find analytical solutions for the optimal times in these cases as well, the expressions becomes more involved and thus we limited our extended analysis in the main text to the simplest case. 

We note that even for X rotations we consider an even number $n+1$ of bangs to build the FATO approximation. This ensures that the function $f(t)$ is odd, yielding a sine Fourier series which is zero at $t=0$ for any bandwidth.\vspace{-12pt}
\subsection{Ultrastrong Driving}
The strong driving occurs when $\theta>\pi/4$. In this case, following \cite{Boscain06} and \cite{Aiello15q}, we find that the TO solution is composed of at most three bangs. It is then easy to find analytical solutions for the times $\{t_i, t_m, t_f\}$, for example by following the construction described in \cite{Piovan12}. 

We can first verify that two bangs are not enough to generate the desired rotations. Defining $\vec v_{\pm,0}$ the vectors corresponding to the Hamiltonians $\ham_{\pm,0}$ and $R=X,Y$ the rotations in SO(3) corresponding to $\sx, \sy$, we need to verify whether $\vec v_i\cdot \vec v_j=\vec v_i\cdot R\cdot \vec v_j$. However $\vec v_+\cdot \vec v_m=\cos(2\theta)$, $\vec v_\pm\cdot \vec v_0=\cos(\theta)$, while $\vec v_0\cdot R\cdot \vec v_\pm=-\cos(\theta)$,  $\vec v_+\cdot Y\cdot \vec v_-=-\cos(2\theta)$ and $\vec v_+\cdot X\cdot \vec v_-=-1$.

Similarly, we can easily identify all the allowed three-bang constructions that achieve the desired unitaries. We find that the central bang must be singular ($\Omega=0$) for the $\sy$ gate. Then, since the desired $\sy$ gate cannot have a component along $\sx$ we have to set 
\[\frac{i}2\tr\{(U_\pm U_0U_\mp)\sx\}\!=\!\sin(\theta)\sin\!\left(\!\frac{\omega_0t_2}2\!\right)\sin\!\left[\frac{\omega_0(t_1\!-\!t_3)}2\right]\!=\!0\]
by selecting $t_1^y=t_3^y$ (since $t_2=0$ has already been excluded). Finally, $t_1^y$ and $t_2^y$ can be easily found algebraically, yielding Eq.~(\ref{eq:TimeS}) in the main text.
For the $\sx$ gate, solutions with and without a singular bang are allowed. In both cases the outer rotations are about the same axis, e.g. $U_+U_0U_+$ or $U_+U_-U_+$, and of the same duration, $t_1^x=t_3^x$. 
We can then calculate explicitly the times and compare the two possible solutions to select the time-optimal one. We find that the shortest evolution is obtained by discarding the singular solution, resulting in the times in Eq.~(\ref{eq:TimeS}).
\vspace{-14pt}
\subsection{Time-optimal control of Two Qubits}
Extending the results of BB TO control to more than one qubit is generally difficult, but results have been found for some particular cases~\cite{Assemat10,Romano14,Romano15,Ashhab12}. In particular, it has been found~\cite{Romano15} that for two qubits with opposite drift terms and under the same control,\vspace{-4pt}
\begin{equation}\ham_\pm=\frac{\omega_0}2(\sz^1-\sz^2)\pm\frac\Omega2(\sx^1+\sx^2)\end{equation}
the TO problem can be solved simultaneously. In particular, for $\pi$ rotations, we recover the same solutions found for one qubit. Then we could repeat the analysis performed for FATO control  of one qubit; the fidelity is simply the square of the fidelity found for one qubit.

We note that while these analytical results are restricted to particular cases, they could be at the basis of numerical searches in more complex situations. For example, knowing the control function and required bandwidth for two non-interacting qubits could be used as initial guess for numerical searches of the control profile for two \textit{interacting} qubits.

Bang-bang control can as well be used to achieve two-qubit gates. Time optimal solutions for the steering of two qubits with Hamiltonian $\ham=J\sz^1\sz^2/2$ were indeed found under the assumption of delta pulses (zero-duration pulses at infinite driving power).  While the control solutions obtained when relaxing these assumptions might not be time-optimal, we can still aim to preserve the optimal time and look for the ensuing drop in fidelity. In the main text we considered an exemplary gate, the SWAP gate between two qubits (swapping their states):
\begin{equation}U_{swap}=\left[\begin{array}{cccc}
1&0&0&0\\
0&0&1&0\\
0&1&0&0\\
0&0&0&1	
\end{array}\right]
\end{equation}
\begin{figure}[t]\centering
\includegraphics[width=0.43\textwidth]{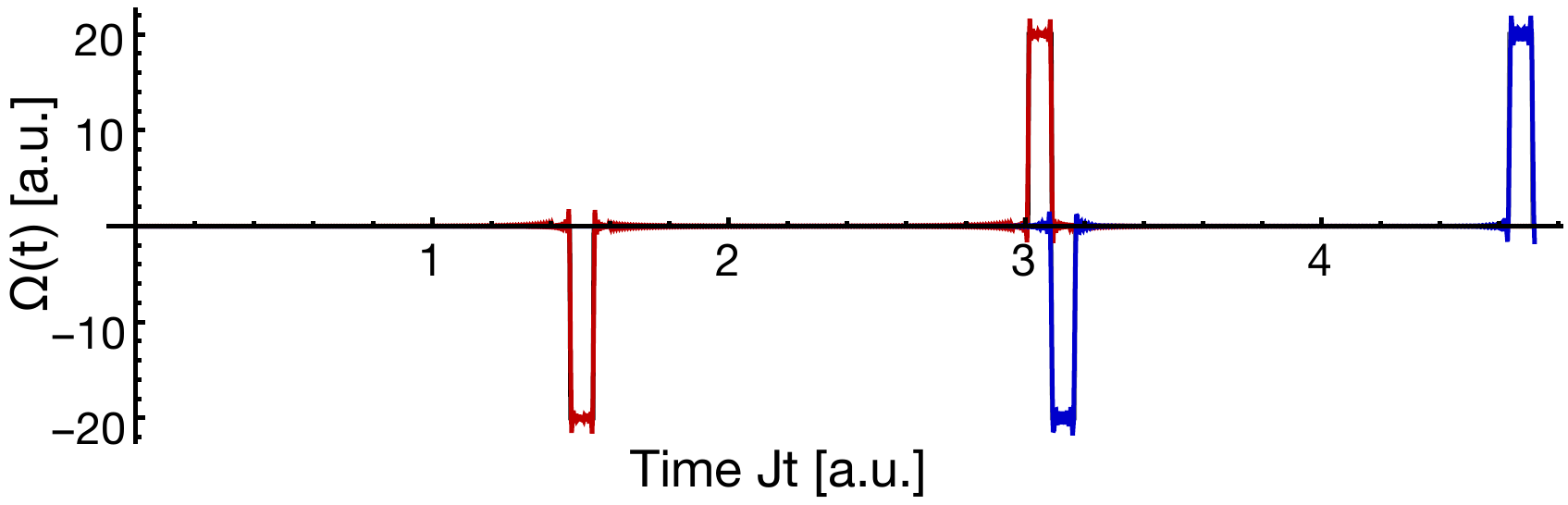}  
\caption{
\textit{SWAP Gate}. We show the control profile for realizing a SWAP gate with the FATO control. In red is the driving along the $x$ axis and in blue along the $y$ axis. For the bandwidth considered here ($\Delta\omega=400J$) the FATO control completely masks the BB control (solid black lines).}  
  \label{fig:Waveform2qb}\vspace{-12pt}
\end{figure}
As shown in~\cite{Khaneja01}, this gate can be obtained in a time $t=3\pi/2J$ by applying instantaneous $\pi/2$ pulses about $x$ and $y$. This ideal control can be in practice replaced by finite-duration (rectangular) pulses, to account for finite control strength. In turn rectangular pulses can be replaced by FATO driving to take into account the control finite bandwidth.
Figure~(\ref{fig:Waveform2qb}) shows the control sequence we implemented to analyze the fidelity behavior of FATO for two qubit control, as shown in Fig.~(\ref{fig:Fid2qb}).

\section{Fidelity}\label{sec:appendixB}
Here we provide further details on the calculations of the fidelity.
Due to FATO control, the system evolves under the Hamiltonian $\ham\!=\!\ham_{id}(t)+\ham_R(t)$, with 
\begin{equation}\begin{array}{l}
\ham_{id}(t)\!=\!\omega_0/2[\sz+f(t)\tan(\theta)\sx]\\
\ham_R(t)\!=\!\omega_0/2\tan(\theta) R_K\!(t)\sx,
\end{array}\end{equation}
 Here $f(t)$ is the BB control function $f(t)=\Omega(t)/\overline\Omega$ and $R_K(t)$ the reminder of its truncated Fourier series,
\begin{equation}
R_K(t)=\sum_{k=K+1}^\infty[s_k\sin({2\pi k t}/T)+ c_k\cos({2\pi k t}/T)].
\label{eq:reminder}
\end{equation}

In order to calculate the infidelity we consider the propagator $U_K$
achieved by implementing a control field according to FATO to order $K$, satisfying 
\begin{equation}
i\dot U_K(t)\!=[\ham_{id}(t)+\ham_R(t)]U_K(t)
\label{eq:U}
\end{equation}
and the propagator due to the ideal BB evolution, $U_{id}(t)$ defined by 
\begin{equation}
i\dot U_{id}(t)=\ham_{id}(t)U_{id}(t)
\label{eq:U}
\end{equation}
The infidelity of the truncated FATO control can be evaluated using the entanglement fidelity~\cite{Nielsen02}
\begin{equation}F=|\tr[U_{id}^\dag(T)U_K(T)]|/2\end{equation}
We can decompose the total propagator $U_K(T)$ as $U_K(T)=U_{id}(T)U_R(T)$, by defining the error propagator $U_R(t)=U_{id}^\dag(t)U_K(t)$. Then the fidelity is simply defined as $F=|\tr[U_{R}(T)]|/2$.

The error propagator can be evaluated by moving to the \textit{toggling} frame defined by the ideal control Hamiltonian. In this frame, the Hamiltonian becomes
\begin{equation}\widetilde\ham_R(t) = U^\dag_{id}(t)\ham_R(t)U_{id}(t)\end{equation}
and the error propagator satisfies  the Schr$\ddot{\textrm{o}}$dinger equation\vspace{-12pt}
\begin{equation}i\dot{\widetilde U}_R(t)=\widetilde\ham_R(t)\widetilde U_R(t)\end{equation}
We can approximate $U_R$ with a first-order Magnus expansion given by the effective Hamiltonian 
\begin{equation}\overline\ham_R=\int_0^T \widetilde\ham_R(t')dt'\end{equation}
Consider for example the weak driving regime. The contribution to $\overline\ham_R$ from each bang is given by the integral over the interval $[t_j,t_{j+1}]$ of 
\begin{equation}
\widetilde\ham_j\!=\!\omega_0\!\tan\!(\theta)R_K\!(t)U_{id}^\dag(t_j) [e^{i\ham_\pm t}\sx e^{-i\ham_\pm t}]U_{id}(t_j)
\label{eq:Ham_j}
\end{equation}
Each pair of ideal propagators $U_-U_+$ creates a rotation $e^{-2i\sy\theta}$. 
Since the angle $\theta$ is small for weak driving, we can approximate this expression by ignoring the time evolution due to $\ham_{id}$ \textit{during} the $j^{th}$ time interval and only considering its effects stroboscopically. Then Eq.~(\ref{eq:Ham_j}) reduces to
\begin{eqnarray}
&&\widetilde\ham_j=\frac{\omega_0}2\tan(\theta)R_K(t)U_{id}(t_{j+1})\sx U_{id}(t_{j+1})\\
&&\qquad = \frac{\omega_0}2\tan(\theta)R_K(t)[(-1)^j\cos(2j\theta)\sx+\sin(2j\theta)\sz]\nonumber
\label{eq:Ham_japp}
\end{eqnarray}
Note that the sign of the $\sx$ terms follows the same pattern as the BB function $f(t)$. Setting $j(t)=\lceil t/T\rceil$, we then need to evaluate the integrals
\begin{equation}\frac2T\int_0^T f(t)\cos[2j(t)\theta]\sin(2\pi kt/T)dt=s_k\frac\theta\pi \end{equation}
and\vspace{-6pt}
\begin{equation}\frac2T\int_0^T \sin[2j(t)\theta]\sin(2\pi kt/T)dt=-s_k\frac\theta\pi\tan(\theta)\end{equation}
We thus obtain the average Hamiltonian \vspace{-12pt}

\begin{equation}
\overline\ham_R=\frac{\omega_0}2\tan(\theta)\frac\theta\pi\sum_{k=K+1}^\infty\!\!\!s_k^2\ [\sx-\tan(\theta)\sz],\vspace{-4pt}
\end{equation}
where we recognize the mean error of the truncated Fourier series, $\mathcal E_k=\frac12\sum_{k=K+1}^\infty\!s_k^2$. \\
The fidelity is then given by $F=\cos(\|\overline\ham_R\|T)$:
\begin{eqnarray}
&&F=\cos\left(\frac{\theta}{\pi}\frac{\omega_0T}{2\cos(\theta)} \tan(\theta)\mathcal E_K \|\cos(\theta)\sx-\sin(\theta)\sz\|\right)\nonumber\\
&&\quad=\cos\left(\frac\pi2\tan(\theta)\mathcal E_K\right).
\end{eqnarray}

A similar calculation can be done for the ultrastrong driving case. However in that case each ``bang'' has a long duration, thus we need to start from Eq.~(\ref{eq:Ham_j}) to find the average Hamiltonian and only approximate or numerical solutions can be found. Still, we find that the solutions depend on the mean Fourier error in a simple way, $F_s^x\approx\cos[2/\pi\sin(\theta)\mathcal E_K]$ and $F_s^y\approx\cos[2/\pi\tan(\theta)\mathcal E_K]$.

%

\end{document}